\documentclass[preprint]{aastex62}
\usepackage{xcolor, soul}
\sethlcolor{green}
\hypersetup{linkcolor=red,citecolor=blue,filecolor=cyan,urlcolor=magenta}

\shorttitle{Seismology of photospheric bright points.}
\shortauthors{Cho et al.}

\begin{document}

\title{Seismological determination of the Alfv\'{e}n speed and plasma-beta in solar photospheric bright points}

\correspondingauthor{Yong-Jae Moon}
\email{moonyj@khu.ac.kr}

\author[0000-0001-7514-8171]{Il-Hyun Cho}
\affiliation{Department of Astronomy and Space Science, Kyung Hee University, Yongin, 446-701, Republic of Korea}

\author[0000-0001-6216-6944]{Yong-Jae Moon}
\affiliation{Department of Astronomy and Space Science, Kyung Hee University, Yongin, 446-701, Republic of Korea}
\affiliation{School of Space Research, Kyung Hee University, Yongin, 446-701, Republic of Korea}

\author[0000-0001-6423-8286]{Valery M. Nakariakov}
\affiliation{School of Space Research, Kyung Hee University, Yongin, 446-701, Republic of Korea}
\affiliation{Centre for Fusion, Space and Astrophysics, Department of Physics, University of Warwick, Coventry CV4 7AL, UK}

\author[0000-0003-1459-3057]{Dae Jung Yu}
\affiliation{School of Space Research, Kyung Hee University, Yongin, 446-701, Republic of Korea}

\author[0000-0001-6412-5556]{Jin-Yi Lee}
\affiliation{Department of Astronomy and Space Science, Kyung Hee University, Yongin, 446-701, Republic of Korea}

\author[0000-0003-1859-0515]{Su-Chan Bong}
\affiliation{Space Science Division, Korea Astronomy and Space Science Institute, Daejeon 305-348, Republic of Korea}
\affiliation{Department of Astronomy and Space Science, University of Science and Technology, Daejeon 305-348, Republic of Korea}

\author{Rok-Soon Kim}
\affiliation{Space Science Division, Korea Astronomy and Space Science Institute, Daejeon 305-348, Republic of Korea}
\affiliation{Department of Astronomy and Space Science, University of Science and Technology, Daejeon 305-348, Republic of Korea}

\author[0000-0003-2161-9606]{Kyung-Suk Cho}
\affiliation{Space Science Division, Korea Astronomy and Space Science Institute, Daejeon 305-348, Republic of Korea}
\affiliation{Department of Astronomy and Space Science, University of Science and Technology, Daejeon 305-348, Republic of Korea}

\author{Yeon-Han Kim}
\affiliation{Space Science Division, Korea Astronomy and Space Science Institute, Daejeon 305-348, Republic of Korea}
\affiliation{Department of Astronomy and Space Science, University of Science and Technology, Daejeon 305-348, Republic of Korea}

\author{Jae-Ok Lee}
\affiliation{Space Science Division, Korea Astronomy and Space Science Institute, Daejeon 305-348, Republic of Korea}

\begin{abstract}
The Alfv\'{e}n speed and plasma beta in photospheric bright points observed by the Broadband Filter Imager (BFI) of the Solar Optical Telescope (SOT) on board the \textit{Hinode} satellite, are estimated seismologically.
The diagnostics is based on the theory of slow magnetoacoustic waves in a non-isothermally stratified photosphere with a uniform vertical magnetic field.
We identify and track bright points in a G-band movie by using the 3D region growing method,
and align them with blue continuum images to derive their brightness temperatures.
From the Fourier power spectra of 118 continuum light curves made in the bright points,
we find that light curves of 91 bright points have oscillations with properties which are significantly different from oscillation in quiet regions,
with the periods ranging 2.2--16.2~min.
We find that the model gives a moderate value of the plasma beta when $\gamma$ lies at around 5/3.
The calculated Alfv\'{e}n speed is 9.68$\pm$2.02~km~s$^{-1}$, ranging in 6.3--17.4~km~s$^{-1}$.
The plasma beta is estimated to be of 0.93$\pm$0.36, ranging in 0.2--1.9.
\end{abstract}


\section{Introduction} \label{sec:intro}
A bright point (BP) is a localized concentration of the solar photospheric magnetic flux that is passively advected and buffeted by granular flows \citep{2011A&A...531L...9M}, formed at an inter-granular lane.
A typical BP diameter is 100--300 km whose upper limit is smaller than the width of the darker lane \citep{2004A&A...422L..63W}.
The brightness of BP comes from a deeper and hotter layer due to the presence of the magnetic pressure \citep{1976SoPh...50..269S, 1993A&A...277..639S, 1998ApJ...494..851M, 2004A&A...422..693M}.
Depending on their positions, chains of BPs are observed as filigree, faculae, or network bright points \citep{1981SoPh...69....9W, 2004ApJ...610L.137C}.
BPs cover 0.19$-$0.88\% of the solar surface and significantly contribute to the variation of the total solar irradiance \citep{2012A&A...539A...6B}.
BPs are associated with roots of spicules \citep{1995ApJ...450..411S, 2012ApJ...744L...5J}. The BP geometry suggests that they could act as magnetohydrodynamic (MHD) wave guides
\citep[e.g.,][]{2008JKAS...41..173K, 2011ApJ...736L..24O, 2012ApJ...750...51K, 2012ApJ...744L...5J, 2017NatSR...743147S, 2018ApJ...854....9S},
and thus may contribute to high atmospheric heating.

\citet{2001ApJ...553..449B} demonstrated that G-band BPs are concentrations of the magnetic flux
associated with down-flows \citep[e.g.,][]{2009A&A...504..583F}.
The inclination angle of bright grains in the photospheric network near the solar disk center is found to be lower than 10$^{\circ}$ \citep{1994ApJ...424.1014S}.
The difference of diameters of BPs in the co-aligned G-band and Ca II images is a few tens of kilometer \citep{2011CEAB...35...29K, 2017ApJ...851...42X},
which is much smaller than the difference of the formation heights of these lines.
Thus, BPs are modeled as thin vertical magnetic flux tubes of a kG field strength.
Such small magnetic structures are suggested to act as effective waveguides for fast and slow magnetoacoustic waves.

Various MHD oscillations have been observed in the solar atmosphere \citep[e.g.,][]{2016SSRv..200...75N}.
In particular, recent studies have revealed that magnetoacoustic oscillations of magnetic flux tubes are associated with many dynamical phenomena.
For example, \citet{2018NatPh..14..480G} observed local temperature enhancements in a sunspot,
and interpreted it as the indication of the dissipation of Alfv\'{e}n waves driven by upwardly propagating magnetoacoustic oscillations.
\citet{2014ApJ...789..108C} reported that fibrils in the superpenumbra of a sunspot are powered by sunspot oscillations.
\citet{2009A&A...505..791S} established that 3-min quasi-periodic pulsations in a flare are triggered by a slow wave leaking from a sunspot.
\citet{2012ApJ...744L...5J} provided evidence for the conversion of longitudinal modes to transverse by analyzing
the intensity variations of BPs with displacement oscillations of the associated spicules.
\citet{2018ApJ...854....9S} verified this scenario by performing an one-dimensional numerical simulation.

\citet{2009ApJ...702.1443F} reported oscillations with the periods ranging from 4 to 9~min in nine intergranular magnetic structures
observed with \textit{Hinode} \citep{2007SoPh..243....3K}/Solar Optical Telescope \citep[SOT,][]{2008SoPh..249..167T},
and performed seismology assuming that both longitudinal and transverse modes had been detected.
\citet{2012ApJ...746..183J} observed intensity oscillations with periods in the range of 110--600~s in the G-band and continuum obtained from
the Rapid Oscillations in the Solar Atmosphere (ROSA) instrument \citep{2010SoPh..261..363J} available on \textit{Dunn Solar Telescope},
and found that 73\% of BPs and 71\% of non-BPs display upwardly propagating wave phenomena.
The oscillating frequencies in BPs might be different from those in non-BPs \citep[see Figure 4,][]{2012ApJ...746..183J},
possibly due to the presence of magnetic field that decreases the plasma beta.

In this Letter, we apply seismological diagnostics to continuum intensity oscillations in BPs observed near the solar disk center (81$\arcsec$, 21$\arcsec$).
Our technique is based on the theory of slow magnetoacoustic waves in a vertical magnetic flux embedded in a non-isothermal stratified atmosphere \citep{2006RSPTA.364..447R}.
The results allow us to estimate the  Alfv\'{e}n speed and plasma beta, that otherwise require simultaneous measurements of the density and magnetic field strength as well as the temperature
at the same location in the photosphere.
The following Section provides a description of the data and tracking method.
In Section 3, the Alfv\'{e}n speed and plasma beta of BPs are obtained by the seismic inversion.
Finally, we summarize our results.

\begin{figure}
\includegraphics[scale=0.8]{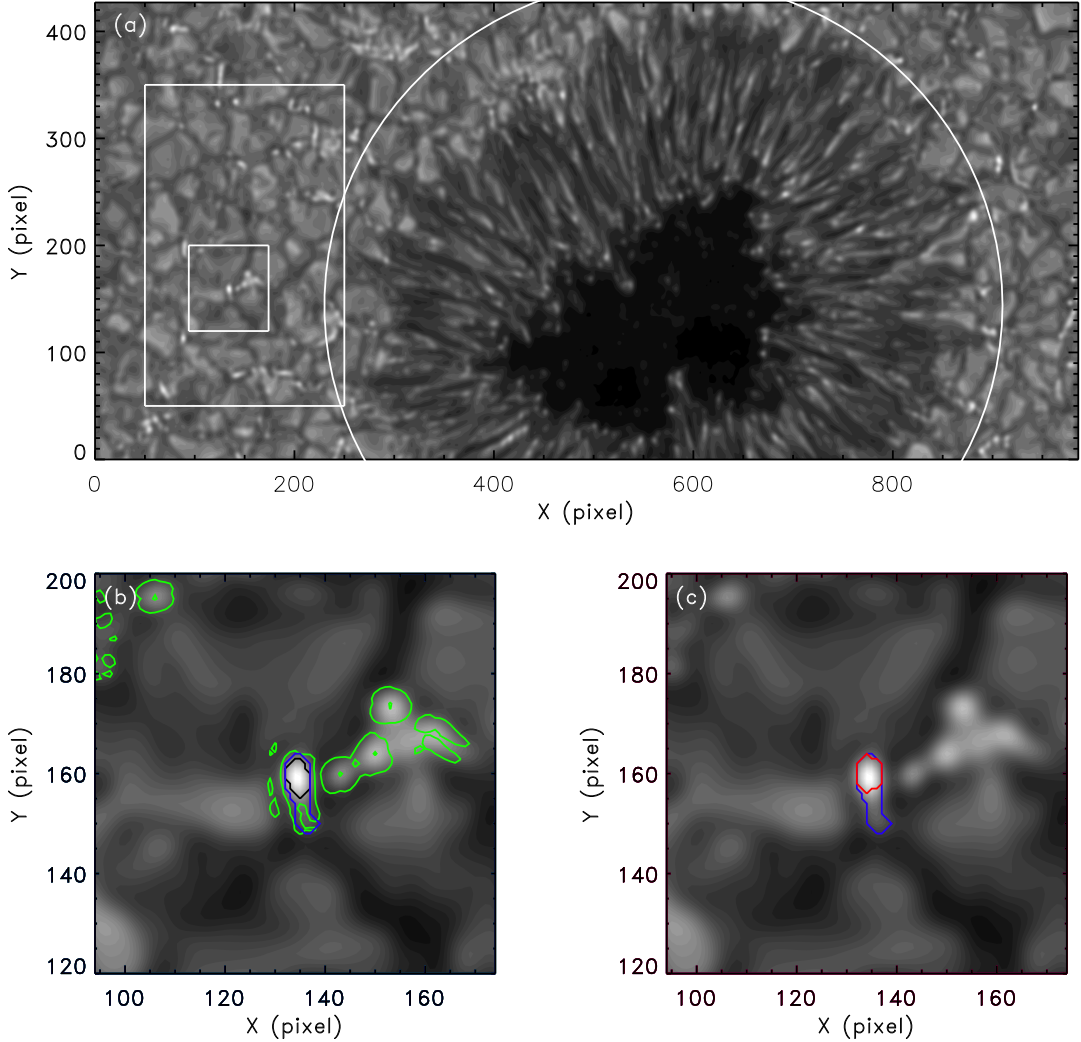}
\caption{
A snapshot of a deconvolved G-band image (a) with the smaller box including a BP.
The white circle is the boundary used to exclude umbral dots and penumbral roots.
The larger box indicates the region used for the measurement of the background intensity.
The smaller box shows an example of a local region with the size of 81$\times$81 including a BP.
The black contour (b) is the boundary of BP determined by the 3D region growing method with the threshold intensity of 0.05.
The green contour (b) represents the region where the square root of the intensity gradient exceeds 3$\sigma$ in the local box,
whose intensity average provides a new threshold intensity appropriate for the local box.
The blue contour (b, c) indicates the adjusted boundary of the BP thresholded by the average intensity of the high gradient region defined by the green contour.
The red contour (c) indicates the final boundary of the BP excluding the merging or separating bright patches close to the main BP.
}
\label{Figure1}
\end{figure}

\section{Data and Method} \label{sec:dm}

The G-band and continuum movies are observed by \textit{Hinode}/SOT BFI during 01:36--02:31~{UT} on 2007-Mar-01 with the time cadence of 6.48~s.
The field-of-view is 56$\arcsec\times$28$\arcsec$ located near the solar disk center (81$\arcsec$, 21$\arcsec$) with the grid spacing of 0.05448$\arcsec$.

We deconvolve the stray lights in both movies by using the point spread functions (PSFs) measured during the Mercury transit on 2006-Nov-8 \citep{2009A&A...501L..19M}.
A first guess ($I_\mathrm{1}$) is defined by the original image ($I_\mathrm{0}$) multiplied by a convolution of a normalized image with the PSF ($P$),
where the normalized one represents the original image divided by a convolution of the original image ($I_\mathrm{0} / (I_\mathrm{0} \ast P)$).
In other words, the initial guess can be expressed by $I_\mathrm{1} = I_\mathrm{0}\times( (I_\mathrm{0}/(I_\mathrm{0} \ast P)) \ast P)$,
where $\ast$ is the convolution operator \citep[e.g.,][]{2012ApJ...752..109L}.
The task is iteratively performed until $I_\mathrm{i}$ converges.
The linear fitting coefficients for $I_\mathrm{i+1}$ as a function of $I_i$ are calculated after each deconvolution.
If the first order coefficient deviates less than 10$^{-3}$ from the unity, we take $I_\mathrm{i}$ as an approximation of the unconvolved image.
After that, we align the G-band and continuum movies separately, and adjust small offset between the G-band and continuum images by performing the alignment once again.

\begin{figure}
\includegraphics[angle=90, scale=0.65]{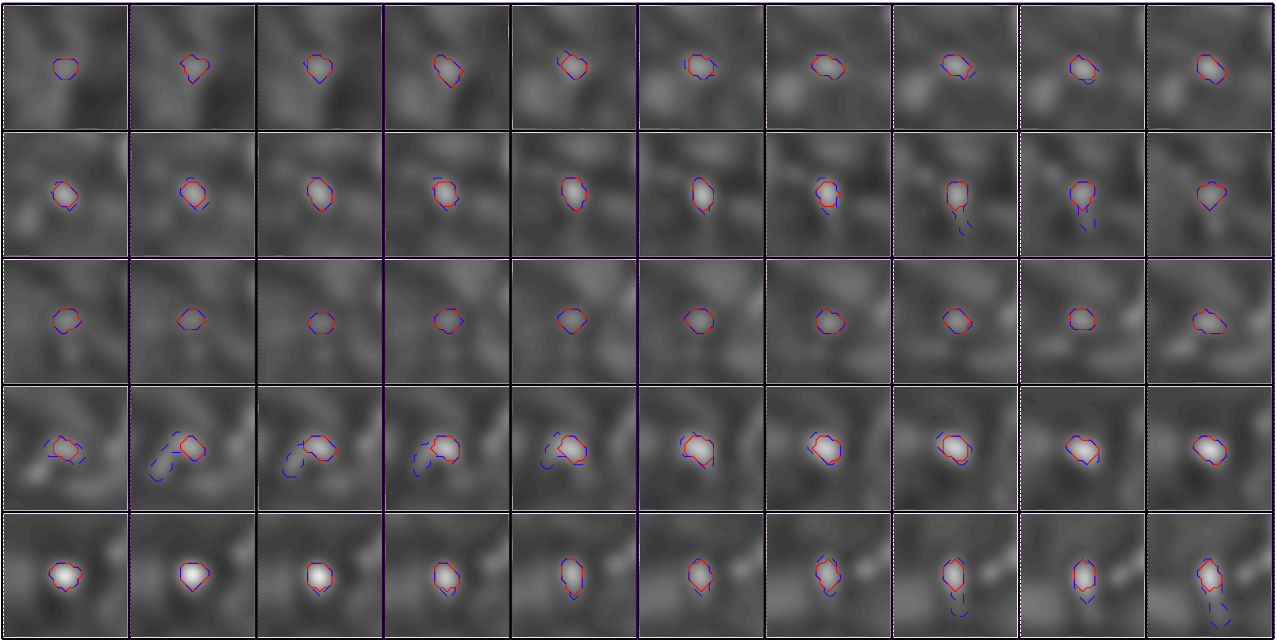}
\caption{
Evolution of a BP with the time step of 19.44 s. The blue and red colors indicate the same as in Figure \ref{Figure1}.
}
\label{Figure2}
\end{figure}

Each frame of the G-band movies is divided by the background intensity which is calculated from the average intensity of the large white box in Figure \ref{Figure1}a.
Then we subtract a smoothed image with 3$\times$3 kernel with uniform weights to amplify the bright signals \citep{2010ApJ...725L.101A}.
To define BPs in the G-band, (1) we apply the 3D region growing method \citep[e.g.,][]{Xiong2016CAA40..540} by using the intensity contrast threshold of 0.05
starting from a certain position and time of a maximum contrast in the G-band movie,
which gives an initial candidate of the pixel positions (black contour in Figure \ref{Figure1}b) of the BP as a function of time.
The threshold 0.05 (5\%) represents bright targets such as BP, umbral dots, and penumbral bright roots.
(2) We define a sub-movie of the 81$\times$81 pixel size, whose center is the average position of the BP (the smaller white box in Figure \ref{Figure1}b) and then
the threshold contrast is adjusted by taking the average contrast of high gradient region where the square root of the gradient is larger than 3 sigma in the local box
(green contours in Figure \ref{Figure1}b) \citep[e.g.,][]{2014SoPh..289.1143T, 2015ApJ...811...49C}.
As a result, we obtain the boundary of BP denoted with the blue contour as shown in the lower panels.
(3) To exclude bright patches too close to the main BP, which are merging or being separated ones,
we check the distribution of contrast as a function of distance in the ascending order from the maximum contrast pixel.
From a certain position, if the contrast starts to increase, we exclude those pixels located outside this position.
As a result, we obtain the BP boundary as noted with the red contour in Figure \ref{Figure1}c.
The maximum allowed deviation of the distance between locations of the maximum contrast in two consecutive images is $\sqrt{2}$ pixels.
(4) Finally, we fill in the obtained pixel address with a null value, and repeat steps (1)--(3) for a next position of the maximum contrast, until the maximum value becomes lower than 0.05.
Snapshots of an example BP evolution are presented in Figure \ref{Figure2}, showing that merging or separating patches close to the main BP are well distinguished.

To extract a BP light curve, the positions of G-band BP is overlapped on the aligned blue continuum image.
In this way, we defined approximately 8000 light curves of isolated bright patches.
Among them, we neglect samples inside the white circle as shown in Figure~\ref{Figure1}a to exclude umbral dots and penumbral roots,
and obtain 118 BPs whose lifetime is longer than 5 min.
The median size of 118 BPs is $\sim$20 pixels.
The median length of BP light curves is 67 frames.

\begin{figure}
\includegraphics[angle=90, scale=0.65]{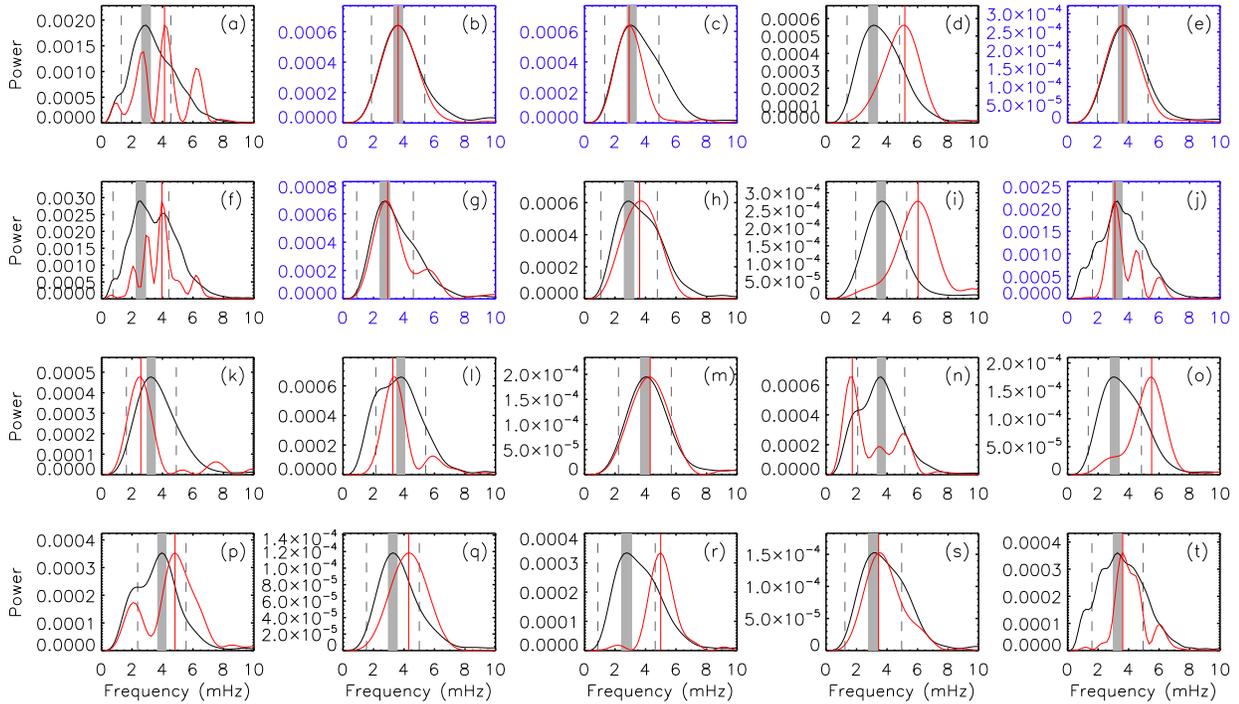}
\caption{
Power spectra of 20 BPs (red) within the continuum and the corresponding mean spectrum of the quiet region (black).
The vertical red line indicates peak frequencies of BPs ($f_\mathrm{BP}$).
The black dashed line indicates $f_\mathrm{non-BP}\pm\sigma_\mathrm{f}$.
The gray region covers $f_\mathrm{non-BP}\pm2.576\sigma_\mathrm{f}/\sqrt{N}$, where $N$ represents the number of light curves in the non-BP region.
Note that the $t$-statistics is defined by $t=(f_\mathrm{BP}-f_\mathrm{non-BP})/(\sigma_\mathrm{f}/\sqrt{N})$.
}
\label{Figure3}
\end{figure}

Any given time series of BPs has a significantly high spectral power near 30--90~s, which are difficult to filter out.
Thus, for a given BP, we take 7 intensity time series by taking samples at every 7 points starting from the 1st to 7th frame from the original light curve.
Then we use 2nd order derivatives of the signals to minimize the effect of the growth and decay of the BP, and subtract linear trends determined by best-fitting.
As a result, 7 stationary time series are prepared whose Nyquist frequency is equal to (90.72~s$^{-1}$).
These re-sampled time series are averaged to obtain a BP light curve.
From this, we obtain the power spectrum for a given BP.

As the continuum formation height of a BP is depressed comparing to the ambient by $\sim$100~km \citep{2010A&A...515A.107S},
the plasma density in the BP could be much larger than in the quiet region.
Thus, a direct comparison of the oscillation spectra in a BP and non-BP is not allowed.
Instead, due to the presence of the magnetic field in a BP, it is likely that the oscillating frequency would be different from those in quiet regions.
Therefore, we perform the student-$t$ test whether the peak frequency of a given BP ($f_\mathrm{BP}$) differs from the sample population of peak frequencies seen in the surrounding quiet region (non-BP region).
A non-BP light curve is obtained by averaging the intensities of a surrounding region whose size is the same with the BP (typically $\sim$20 pixels), within the same local box.
In this way, about 250 non-overlaping light curves and their power spectra are defined for a given BP.
From the spectra, we determine the mean spectrum, which defines the mean ($f_\mathrm{non-BP}$) and standard deviation ($\sigma_\mathrm{f}$) of the peak frequencies of the sample population.
In Figure~\ref{Figure3}, power spectra of 20 BPs and the corresponding peak frequencies as well as those of quiet regions are presented.
In most of BPs, their peak frequencies are statistically different from those of non-BPs with 99\% level of confidence ($|f_\mathrm{BP}-f_\mathrm{non-BP}|>2.576\sigma_\mathrm{f}/\sqrt{N}$),
where $N$ represents the number of light curves in a non-BP region.
In a few samples as in Figure~\ref{Figure3}b, \ref{Figure3}c, \ref{Figure3}e, \ref{Figure3}g, \ref{Figure3}j,
two peak frequencies are statistically the same with each other, that is in contradiction to the presence of magnetic field.
We find that the peak frequencies in 91 BPs out of 118 are significantly different from those in non-BPs ($p-$value$<$0.01).
The peak periods of 91 BPs range 2.4--16.2~~min.

Because the power spectra often show asymmetric lobes or multiple peaks around the peak frequency,
we use the weighted frequency defined as $\langle f \rangle =\Sigma (P(f) f) /\Sigma P(f)$,
where $P$ and $f$ represent the power and frequency, respectively.
The summation is performed over all discrete frequencies.
The weighted frequencies for 91 BPs ranges 2.8--7.0~mHz (4.24$\pm$0.78~mHz).

\section{Results} \label{sec:results}
We apply a seismological diagnostics based on the intensity oscillation of BPs,
using the relation between the angular cutoff frequency ($\Omega$), sound speed ($c_\mathrm{S}$), and tube speed ($c_\mathrm{T}$).
The temperature is determined from the equation
$I(e^{hc/\lambda k_\mathrm{B} T_\mathrm{B,0}}-1)=e^{hc/\lambda k_\mathrm{B} T_\mathrm{Q,0}}-1$,
where $I$, $h$, $c$, $\lambda$, $k_\mathrm{B}$, $T_\mathrm{B,0}$, and $T_\mathrm{Q,0}$ are
the intensity ratio, Planck's constant, speed of light, wavelength of blue continuum (4504.5~\AA), Boltzmann's constant, the average temperature at the BP,
and the temperature of the quiet Sun (6,000 K), respectively.
The sound speed is defined as $c_\mathrm{S}=\sqrt{\gamma k_\mathrm{B} T_\mathrm{B,0} / (\mu m_\mathrm{H})}$,
where $\gamma$, $\mu$, and $m_\mathrm{H}$ indicate the adiabatic index, mean molecular weight, and mass of hydrogen, respectively.

The angular cutoff frequency of the slow mode wave in a non-isothermal stratified atmosphere with a uniform vertical magnetic field \citep{2006RSPTA.364..447R} can be expressed as
\begin{equation} \label{eq1}
\Omega^2 = \frac{3\gamma g}{4\Lambda_\mathrm{p}}X^3 - \left(\frac{g}{2\Lambda_\mathrm{p}} + \frac{\gamma g}{\Lambda_\mathrm{p}}\right)X^2 + \frac{g}{\Lambda_\mathrm{p}}X,
\end{equation}
where $\Omega$, $\gamma$, and $g$, are $2\pi f_\mathrm{cutoff}$, adiabatic index, and gravitational acceleration (0.274 km s$^{-2}$), respectively;
$X = c_\mathrm{T}^2/c_\mathrm{S}^2$,  and $\Lambda_\mathrm{p} = c_\mathrm{S}^2/(\gamma g)$ is the pressure scale height.
The cutoff frequency represents a resonant frequency of the medium, but also acts as a high-pass filter.
Thus it is likely to be lower than an observed peak frequency,
which is independent of whether the oscillating medium behaves as a resonator or a filter.
In previous studies, the cutoff frequency has been determined by taking a frequency at the median power in the azimuthally averaged power map of sunspot oscillations \citep{2014A&A...561A..19Y},
or it has been determined from the lowest frequency with a power of three times above the noise level in the frequency range 4--9~mHz \citep{2012ApJ...746..119R}.
In this study, the cutoff frequency is determined by scaling the weighted frequency with the coefficient of 0.53 (i.e., $f_\mathrm{cutoff}$=0.53$\langle f \rangle$),
which is similar to the value of 0.55 used in the previous study \citep{2017ApJ...837L..11C}.
This value is given after a simulation with the atmospheric parameters corresponding to the height of the continuum formation.
If we use the temperature, density, and magnetic field strength of 6,000~K, 10$^{-3.8}$~kg~m$^{-3}$, and 1,300~G, respectively \citep{2010A&A...515A.107S},
then Eq.~(\ref{eq1}) gives the cutoff frequency of 2.23~mHz.
As a result, the scaling coefficient corresponding to the observed weighted frequency becomes 0.53 (= 2.23/4.24).

\begin{figure}
\includegraphics[scale=0.9]{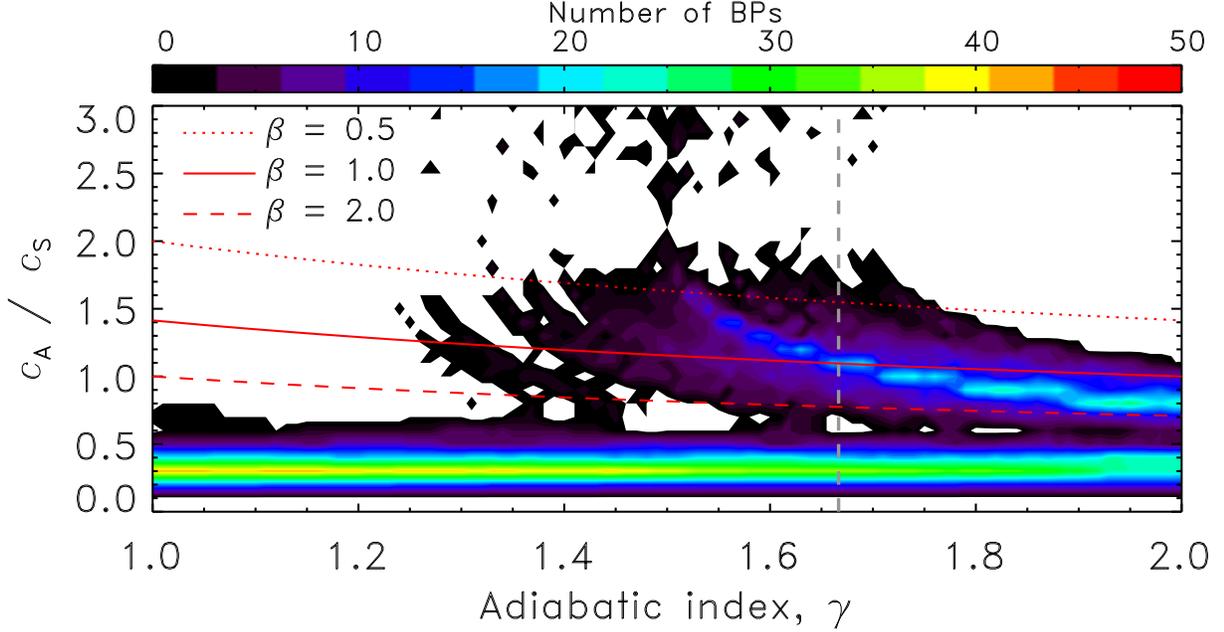}
\caption{
The number of BPs which give real solutions as a function of $c_\mathrm{A}/c_\mathrm{S}$ and $\gamma$.
The gray vertical-dashed line indicates $\gamma$=5/3.
The dotted, solid, and dashed lines indicate $c_\mathrm{A}/c_\mathrm{S}$ ($=\sqrt{2/\left(\gamma\beta\right)}$) for $\beta$=0.5, 1.0, and 2.0, respectively.
}
\label{Figure4}
\end{figure}

Eq.~(\ref{eq1}) is a 3rd order polynomial with respect to $X$ and has a form $\Omega = \Omega(\Lambda_\mathrm{p}, X)$.
Three solutions with complex numbers are obtained for a given BP.
Among them, we select the solutions giving $c_\mathrm{A}>0$, which is satisfied by the condition with  $Im\left\{ X\right\} =$ 0 and 0 $<Re \left\{X \right\}<$ 1.
Because the temperature and the cutoff frequency are obtained observationally,
one can determine the tube speed ($c_\mathrm{T} = c_\mathrm{S}\sqrt{X}$),
Alfv\'{e}n speed ($c_\mathrm{A} = \sqrt{c_\mathrm{T}^2c_\mathrm{S}^2/(c_\mathrm{S}^2 - c_\mathrm{T}^2)} = c_\mathrm{S}\sqrt{X/(1-X)}$),
and plasma beta ($\beta = 2c_\mathrm{S}^2/(\gamma c_\mathrm{A}^2) = 2(1/X-1)/\gamma$) which correspond to a real value of $X$.
The BP brightness can be changed by ambient motions of granules and may be affected by radiation from the lateral side walls.
Therefore, there is a possibility of deviation for adiabatic index from $5/3$.
In Figure~\ref{Figure4}, we draw the number distribution of BPs which give real solutions as a function of $c_\mathrm{A}/c_\mathrm{S}$ and $\gamma$,
to see which value of $\gamma$ can give a moderate value of beta, $\beta \sim 1.0$.
It is found that a moderate beta solution can be allowed when the adiabatic index is around at $\gamma \sim$ 5/3.
Hence, we use the conventional value, $\gamma =$ 5/3.
Eq.~(\ref{eq1}) gives two real solutions of $X$ for a given BP with $\gamma =$ 5/3, and we use the higher one.
The lower one gives an extremely high plasma beta which is in contradiction to the fact that the BP is a vertical magnetic flux tube with a kG field strength.
As a result, we obtain solutions of the moderate plasma beta for 70 BPs out of the 91 significant samples.

\begin{figure}
\includegraphics[angle=90, scale=0.65]{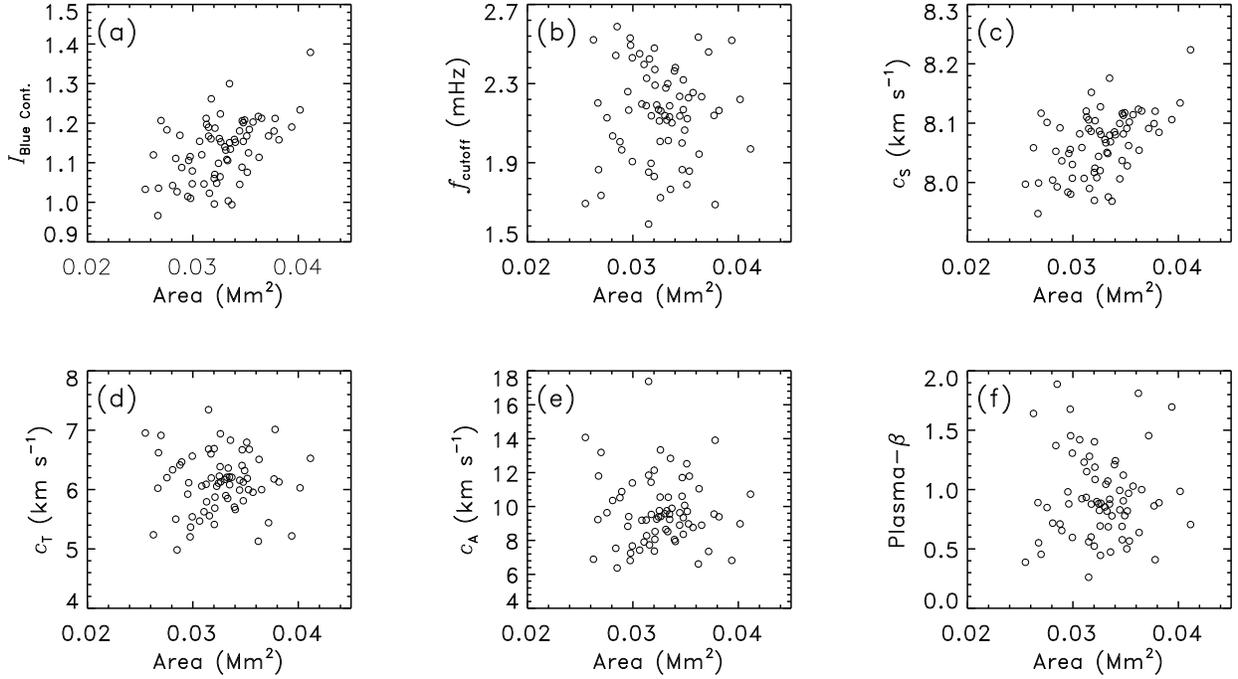}
\caption{
Properties (the mean intensity (a), cutoff frequency (b), sound speed (c), tube speed (d), Alfv\'{e}n speed (e), and plasma-beta (f)) of 70 BPs as a function of area.
}
\label{Figure5}
\end{figure}

Figure~\ref{Figure5} summarises the properties of 70 analysed BP, obtained by the above technique.
The mean intensities of BP light curves are higher than the background intensity.
The correlation coefficient (CC) of the BP brightness and its area is 0.50 with $p-$value$<$0.01,
indicating that larger BP tends to be brighter, which is consistent with the previous finding \citep[e.g.,][]{2010ApJ...725L.101A}.
The cutoff frequency is found to be 2.15$\pm$0.24~mHz, ranging 1.5--2.6~mHz.
The cutoff frequency does not correlate with the BP area (CC = 0.06).
The sound speed is distributed in a narrow range 7.9--8.3~km~s$^{-1}$ (8.07$\pm$0.05~km~s$^{-1}$).
The tube speed ranges in 4.9--7.4~km~s$^{-1}$ (6.11$\pm$0.50~km~s$^{-1}$), and the Alfv\'{e}n speed in 6.3--17.4~km~s$^{-1}$ (9.68$\pm$2.02~km~s$^{-1}$).
The plasma beta is found to range in 0.2--1.9, with the calculated average value 0.93$\pm$0.38 which is comparable to the value of 0.86$\pm$0.07 observed in pores \citep{2017ApJ...837L..11C}.
The CCs for the tube speed, Alfv\'{e}n speed, and plasma with the area of BP are 0.02, -0.04, -0.01, respectively.
The significance level for the latter four CCs are approximately lower than 63\%, 87\%, 75\%, and 94\%.
Hence, the null hypothesis of a zero correlation cannot be rejected,
implying that the cutoff frequency, tube speed, Alfv\'{e}n speed, and plasma beta do not correlated with the BP area.

\section{Summary and Discussion} \label{sec:sd}
We seismologically estimated MHD properties of BPs observed in continuum by their oscillations.
Boundaries of BPs are determined by the 3D region growing technique with an exclusion of merging or separating bright patches during the evolution.
In 118 BPs whose life time is longer than 5 min, we find that 70 oscillations are distinguished from surrounding quiet regions.
The obtained peak period ranges in 2.2--16.2 min.
The calculated Alfv\'{e}n speed is distributed from 6.3 to 17.4~km~s$^{-1}$ which is comparable to the value of the sound speed.
The plasma beta ranges 0.2--1.9.
These values are generally consistent with the BP model developed in \citet{2010A&A...515A.107S},
which shows a moderate beta slightly above the height where $c_\mathrm{A}/c_\mathrm{S}$ is the unity (i.e., $\beta = 1.2$).
Thus, slow magnetoacoustic waves in a non-isothermal and stratified atmosphere with a uniform vertical magnetic field provide a proper interpretation for the observed continuum oscillations of BPs.

The Alfv\'{e}n speed and plasma beta do not show correlations with the BP area.
Using the obtained results, the plasma density in a BP could be derived when high time-cadence observations of the magnetic field become available.
The developed technique could be applied to a slightly larger magnetic structure such as tiny pores \citep{2010ApJ...723..440C},
which may help to improve our understanding of physical properties of solar atmospheric magnetic flux tubes of various sizes.
As shown in Figure~\ref{Figure4}, the model allows for a high beta plasma for entire range of the adiabatic index.
Recent studies revealed that the adiabatic index in the solar corona are highly deviated from 5/3 \citep{2011ApJ...727L..32V, 2018ApJ...868..149K},
but still difficult to measure it in the lower solar atmosphere.
More precise quantification of the properties of small photospheric magnetic structures may be obtained if the adiabatic index is estimated observationally.

\acknowledgments
We appreciate the constructive comments from an anonymous referee which significantly improved the manuscript.
\textit{Hinode} is a Japanese mission developed and launched by ISAS/JAXA,
with NAOJ as domestic partner and NASA and STFC (UK) as international partners.
It is operated by these agencies in co-operation with ESA and NSC (Norway).
This work is supported by the Basic Science Research Program (BSRP) through the National Research Foundation (NRF)
funded by the Ministry of Education of Korea (NRF-2017R1A6A3A01002649, NRF-2016R1A6A3A11932534, NRF-2016R1A2B4013131),
NRF of Korea Grant funded by the Korean Government (NRF-2013M1A3A3A02042232),
the BK 21 plus program through the NRF funded by the Ministry of Education of Korea,
and the Korea Astronomy and Space Science Institute under the R\&D program, Development of a Solar Coronagraph on the International Space Station
(2017-1-851-00), supervised by the Ministry of Science and ICT.
V.M.N. acknowledges support from the STFC consolidated Grant ST/P000320/1.

\end{document}